\documentclass[sigconf]{acmart}
\AtBeginDocument{%
  \providecommand\BibTeX{{%
    \normalfont B\kern-0.5em{\scshape i\kern-0.25em b}\kern-0.8em\TeX}}}

\setcopyright{acmlicensed}
\copyrightyear{2018}
\acmYear{2018}
\acmDOI{XXXXXXX.XXXXXXX}

\acmConference[Conference acronym 'XX]{Make sure to enter the correct
  conference title from your rights confirmation emai}{June 03--05,
  2018}{Woodstock, NY}
\acmISBN{978-1-4503-XXXX-X/18/06}

\usepackage{algorithm}
\usepackage{algorithmic}
\usepackage{pdfpages}
\usepackage{multirow}
\usepackage[export]{adjustbox} 
\usepackage{bm}
\usepackage{amsmath}
\usepackage{array}
\usepackage{booktabs}
\usepackage{footmisc}

\usepackage{amsthm}
\usepackage{graphicx}
\usepackage{enumitem}

\theoremstyle{definition}

\usepackage{subfigure}
\usepackage{hyperref}
\usepackage{url}
\usepackage{color}
\usepackage{colortbl}
\usepackage{xspace}
\usepackage{tabularx}

\newcommand{\model}{ColdLLM\xspace}

\usepackage{amsmath,amsfonts,bm}

\def\eqref#1{equation~\ref{#1}}
\def\1{\bm{1}}

\def\vc{{\bm{c}}}

\def\ve{{\bm{e}}}
\def\vf{{\bm{f}}}

\def\vs{{\bm{s}}}

\def\mC{{\bm{C}}}

\def\mE{{\bm{E}}}
\def\mF{{\bm{F}}}

\def\mI{{\bm{I}}}

\def\mU{{\bm{U}}}

\DeclareMathAlphabet{\mathsfit}{\encodingdefault}{\sfdefault}{m}{sl}
\SetMathAlphabet{\mathsfit}{bold}{\encodingdefault}{\sfdefault}{bx}{n}

\def\gC{{\mathcal{C}}}

\def\gE{{\mathcal{E}}}
\def\gF{{\mathcal{F}}}

\def\gH{{\mathcal{H}}}
\def\gI{{\mathcal{I}}}

\def\gL{{\mathcal{L}}}

\def\gO{{\mathcal{O}}}

\def\gU{{\mathcal{U}}}

\copyrightyear{2025}
\acmYear{2025}
\setcopyright{acmlicensed}
\acmConference[WSDM '25] {Proceedings of the Eighteenth ACM International Conference on Web Search and Data Mining}{March 10--14, 2025}{Hannover, Germany.}
\acmBooktitle{Proceedings of the Eighteenth ACM International Conference on Web Search and Data Mining (WSDM '25), March 10--14, 2025, Hannover, Germany}
\acmISBN{979-8-4007-1329-3/25/03}
\acmDOI{10.1145/3701551.3703546}
\begin{document}

%%
%% The "title" command has an optional parameter,
%% allowing the author to define a "short title" to be used in page headers.
\title{Large Language Model Simulator for Cold-Start Recommendation}
%%
%% The "author" command and its associated commands are used to define
%% the authors and their affiliations.
%% Of note is the shared affiliation of the first two authors, and the
%% "authornote" and "authornotemark" commands
%% used to denote shared contribution to the research.
\author{Feiran Huang}
\email{huangfr@jnu.edu.cn}
\affiliation{%
  \institution{Jinan University}
  \city{Guangzhou}
  \country{China}
}

\author{Yuanchen Bei}
\affiliation{%
  \institution{Zhejiang University}
  \city{Hangzhou}
  \country{China}}
\email{yuanchenbei@zju.edu.cn}

\author{Zhenghang Yang}
\affiliation{%
  \institution{Jinan University}
  \city{Guangzhou}
  \country{China}
}
\email{yangzhenghang@stu2022.jnu.edu.cn}

\author{Junyi Jiang}
\affiliation{%
 \institution{Jinan University}
 \city{Guangzhou}
 \country{China}}
\email{jjy0116@stu2022.jnu.edu.cn}

\author{Hao Chen}
\authornote{Corresponding author.}
\affiliation{%
  \institution{City University of Macau}
  \city{Macao}
  \country{China}}
\email{sundaychenhao@gmail.com}

\author{Qijie Shen}
\affiliation{%
  \institution{Alibaba Group}
  \city{Hangzhou}
  \country{China}}
\email{qjshenxdu@gmail.com}

\author{Senzhang Wang}
\affiliation{%
  \institution{Central South University}
  \city{Changsha}
  \country{China}}
\email{szwang@csu.edu.cn}

\author{Fakhri Karray}
\affiliation{%
  \institution{Mohamed Bin Zayed University of Artificial Intelligence}
  \city{Abu Dhabi}
  \country{UAE}}
\email{fakhri.karray@mbzuai.ac.ae}

\author{Philip S. Yu}
\affiliation{%
  \institution{University of Illinois Chicago}
  \city{Chicago}
  \country{USA}}
\email{psyu@uic.edu}

%%
%% By default, the full list of authors will be used in the page
%% headers. Often, this list is too long, and will overlap
%% other information printed in the page headers. This command allows
%% the author to define a more concise list
%% of authors' names for this purpose.
\renewcommand{\shortauthors}{Feiran Huang, et al.}

%%
%% The abstract is a short summary of the work to be presented in the
%% article.
\begin{abstract}
Recommending cold items remains a significant challenge in billion-scale online recommendation systems. While warm items benefit from historical user behaviors, cold items rely solely on content features, limiting their recommendation performance and impacting user experience and revenue. Current models generate synthetic behavioral embeddings from content features but fail to address the core issue: the absence of historical behavior data. To tackle this, we introduce the \textbf{LLM Simulator} framework, which leverages large language models to simulate user interactions for cold items, fundamentally addressing the cold-start problem. However, simply using LLM to traverse all users can introduce significant complexity in billion-scale systems. To manage the computational complexity, we propose a \textbf{coupled funnel ColdLLM} framework for online recommendation. ColdLLM efficiently reduces the number of candidate users from billions to hundreds using a trained coupled filter, allowing the LLM to operate efficiently and effectively on the filtered set. Extensive experiments show that ColdLLM significantly surpasses baselines in cold-start recommendations, including Recall and NDCG metrics. A two-week A/B test also validates that ColdLLM can effectively increase the cold-start period GMV.

\end{abstract}

%%
%% The code below is generated by the tool at http://dl.acm.org/ccs.cfm.
%% Please copy and paste the code instead of the example below.
%%


%%
%% Keywords. The author(s) should pick words that accurately describe
%% the work being presented. Separate the keywords with commas.

%% A "teaser" image appears between the author and affiliation
%% information and the body of the document, and typically spans the
%% page.
%\begin{teaserfigure}
%  \includegraphics[width=\textwidth]{sampleteaser}
%  \caption{Seattle Mariners at Spring Training, 2010.}
%  \Description{Enjoying the baseball game from the third-base
%  seats. Ichiro Suzuki preparing to bat.}
%  \label{fig:teaser}
%\end{teaserfigure}

%\received{20 February 2007}
%\received[revised]{12 March 2009}
%\received[accepted]{5 June 2009}

%%
%% This command processes the author and affiliation and title
%% information and builds the first part of the formatted document.
\maketitle

\section{Introduction}
Recommending cold items is essential for modern recommender systems, as a continuous flow of new content is being generated by individuals, companies, and AI~\cite{huang2023aldi,wei2017collaborative,yang2024generalized}. Current large-scale recommender systems rely on historical user-item behaviors to learn user and item embeddings and then use these embeddings for downstream recall~\cite{wang2019ngcf,he2020lightgcn,he2017ncf,zhang2024multi} and CTR prediction tasks~\cite{chen2024macro,zhou2018din,zhou2019dien,xiao2020deep}. However, unlike items with user historical interactions, or ``warm'' items, newly added items, or ``cold" items lack behavior data to train embeddings. This lack of behaviors hinders the effective recommendation of cold items to users, impacting the overall ecosystem and the revenue of the recommender system. It is crucial to provide cold items with an embedding to ensure that these items have promising recommendation performance.

Current models typically use the content feature of cold items to generate \underline{synthetic embeddings}~\cite{zhu2021transfer,zhu2021fairness,zhu2021learning}. In particular, \textbf{generative models} attempt to train a mapping function to ensure that the generated embedding approximates the behavior embedding. Representatively, DeepMusic~\cite{van2013deepmusic} accomplishes this by minimizing the discrepancy between the generated embeddings and the actual behavioral embeddings. Expanding on this concept, GAR~\cite{chen2022gar} employs a generative adversarial approach to ensure that the generated embeddings match the distribution of actual behavioral embeddings. ALDI~\cite{huang2023aldi} adopts actual behavioral embeddings as teachers to transfer their knowledge to the generated embeddings. Another line of models, \textbf{dropout models} further enhance the adaptability of recommendation models by incorporating both the generated embeddings and the behavior embeddings. For example, DropoutNet~\cite{srivastava2014dropout} and Heater~\cite{zhu2020heater} typically drop random behavior embeddings and use the generated embeddings instead of real embeddings during training to improve robustness. CLCRec~\cite{wei2021clcrec} utilizes contrastive learning to enhance the compatibility of the generated embeddings and the behavior embeddings.

\begin{figure}
\centering
\includegraphics[width=\linewidth]{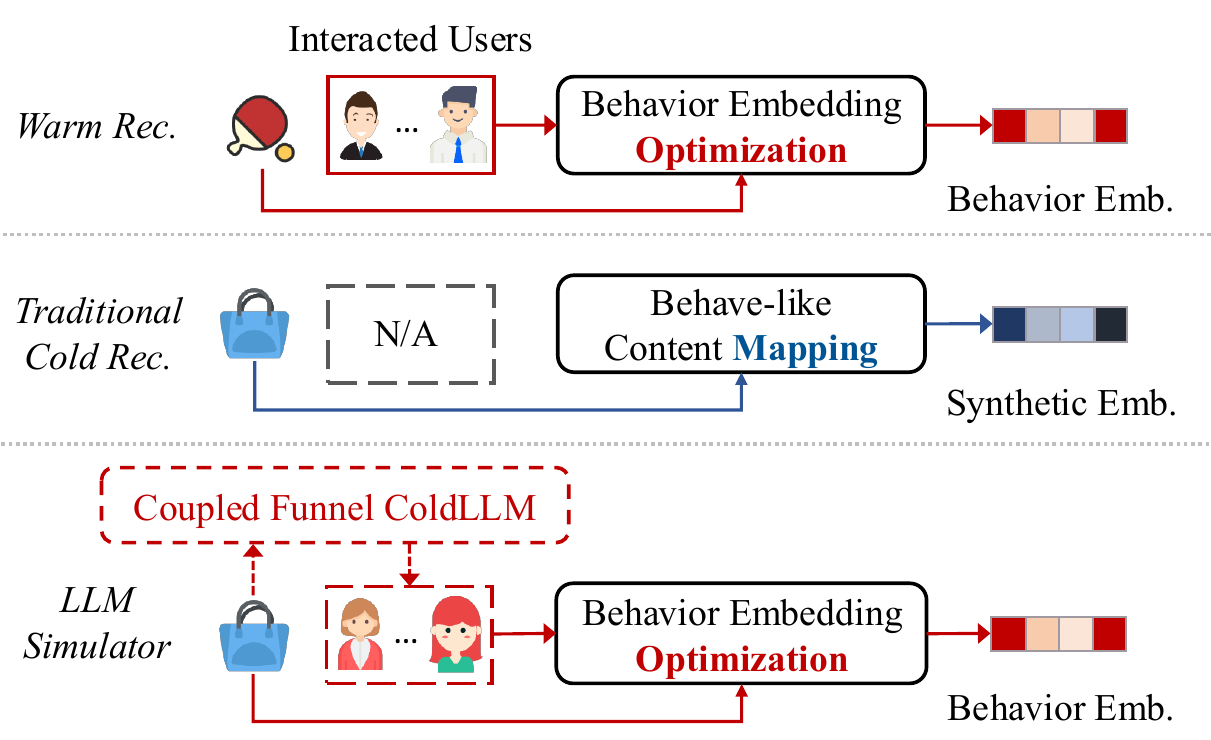}

  \caption{A comparison between traditional item cold-start models and our \model.}
  \label{fig:intro}

\end{figure}

However, existing solutions do not fully address the fundamental problem of cold-start---the lack of behavior data for cold items---which makes cold items inherently different from warm items. As illustrated in Figure~\ref{fig:intro}, this fundamental difference leads to the following three key limitations:
\begin{enumerate}[leftmargin=*]
    \item \textbf{Content-behavior gap:} The synthetic embeddings are still derived from content features. This approach results in a significant discrepancy between these synthetic embeddings and the embeddings learned from actual user behavior and interactions.
    \item \textbf{Suboptimal cold recommendation performance:} Current cold-start models often focus on recommending cold items alongside warm items, without significantly impacting warm items~\cite{huang2023aldi}, less considering improving the recommendation performance for warm items.
    \item \textbf{Conflation of content-based and behavior-based recommendations:} Existing cold-start models typically conduct a mixed recommendation that mixes both content feature embeddings and behavioral embeddings.
\end{enumerate}

Large Language Models (LLMs) show potential in addressing the aforementioned limitations, as they may be capable of understanding user preferences from content features and predicting users' intentions toward items~\cite{lin2023can,wu2024survey}. However, applying LLMs to cold-start item recommendations presents the following challenges:
\begin{enumerate}[leftmargin=*]
    \item \textbf{Simulation of Cold-Start Behavior:} Training an LLM to predict a user's intention towards an item without actual interaction data is a challenge.
    \item \textbf{Efficiency of Simulation:} LLMs face constraints in inference efficiency. Simulating user behaviors for cold items across a large user base incurs substantial computational complexity. 
    \item \textbf{Scalability to Large-Scale Recommendations:} There is an absence of mature frameworks leveraging LLMs to address the cold-start item issue in large-scale recommender systems.
\end{enumerate}

In this paper, we propose a novel \textbf{LLM Simulator} framework to fundamentally address the item cold-start problem. To tackle design challenges, we introduce the tailored structure of the LLM simulator, which includes user context construction, prompt design, and the simulation process. To accelerate the simulation process, we propose the \textbf{ColdLLM} for online recommendation, which efficiently scales down candidate users from billions to hundreds using a trained coupled filter. This filter is trained in conjunction with LLM to support its simulation. We then detail the fine-tuning process of the LLM and the coupled training of the filter model. Lastly, we present the implementation details of ColdLLM on large-scale recommender systems and analyze its complexity.
The key contributions of this study can be summarized as follows:

\begin{itemize}[leftmargin=*]
    \item We formally define the behavior simulation problem and present a novel LLM Simulator framework that fundamentally addresses the cold-start recommendation issue.
    \item We propose a tailored training strategy for the simulator and offer a customized application strategy for the LLM simulator.
    \item We conduct extensive offline experiments, demonstrating that our model outperforms existing solutions by 21.69\% in cold recommendation performance. A two-week A/B test further validates ColdLLM's superiority. 
\end{itemize}

\section{Related Works}

\subsection{Cold-Start Item Recommendation}
Cold-start item recommendations refer to recommending newly occurred items to users. It presents a long-term challenge for recommendation systems due to the lack of behavioral interactions to model these items~\cite{huang2023aldi,wei2021clcrec,liu2024fine,bai2023gorec}.

Currently, one popular type of item cold-start recommendation model typically maps the contents of those cold items to content embeddings and then aligns them with the behavioral embeddings trained on warm items, which can be summarized as the ``\textit{embedding simulation}''.
Among them, one category of methods is the robust co-training models, which aims to align the behavioral embedding of warm items with the content-generated embedding of cold items through co-training with robust strategies~\cite{volkovs2017dropoutnet,zhu2020heater,du2020mtpr,wei2021clcrec,shi2019dropoutmethods1,xu2022dropoutmethods2}.
Another category is the knowledge alignment model, where the goal is to align the embeddings generated from the content of cold items towards the pre-trained behavioral embeddings based on those warm instances~\cite{van2013deepmusic,pan2019metaemb,chen2022gar,huang2023aldi}.
Then, few other efforts have paid attention to the ``\textit{interaction simulation}'', which generates some potential meaningful interactions between cold items and warm users/items~\cite{wang2024mutual,liu2023ucc,cai2023user}.
Representatively, UCC~\cite{liu2023ucc} generates low-uncertainty interactions for cold items that have a similar distribution to warm items with teacher-student consistency learning.
MI-GCN~\cite{wang2024mutual} adopts pair-wise mutual information to generate informative interactions for cold items.

\subsection{Recommendation with LLMs}
Recently, Large Language Models (LLMs) have emerged as a central research focus due to their remarkable ability to understand and generate human-like text, leveraging their pre-trained repository of world knowledge~\cite{gao2024llm,zhao2024recommender,chang2024survey}.

Given the rich tapestry of natural language descriptions inherent in recommender systems, an increasing number of studies are honing in on the potential of LLMs to enhance recommendation capabilities~\cite{li2023llmrec_survey,wei2023llmrec,sanner2023llm4coldstart,bao2023tallrec}.
Representatively, TALLRec~\cite{bao2023tallrec} introduces a novel framework that adeptly integrates LLMs with recommendation tasks through a dual-stage tuning process. TALLRec has been shown to improve recommendation performance and robust cross-domain recommendation tasks. LLMRec~\cite{wei2023llmrec} enhances the recommendation performance by strengthening user-item interaction links, enriching item features, and profiling users from a natural language vantage point. Additionally, the work presented in~\cite{wang2024large} pioneers the use of LLMs to address the cold-start challenge in recommendations, employing the models as data augmenters to generate training signals for cold items through prompting.

In this paper, we focus on leveraging the world knowledge of LLMs and the collaborative filtering capabilities of recommendation models for cold-start item recommendations.

\section{Preliminaries}

\paragraph{\textbf{Notations.}} The user and item sets are denoted as $\gU$ and $\gI$, respectively. In terms of the items, we denote the warm items (items with historical interactions) as $\gI_w$, and the cold items~(items without historical interactions) as $\gI_c$. Let $\gH$ present the set of all the historically interacted user sequences for all the items. Then, each warm item has an interacted user sequence $\vs_i = \{h_{i,1},\cdots, h_{i,|h_i|}\},\ \forall i \in \gI_w$, where $|h_i|$ denotes the number of interaction for item $i$. For a cold item $j$, the interacted user sequence is a null set, namely $\vs_j = \phi$. With the historically interacted user-item pairs, we can learn the behavioral embedding vectors for each user and warm item, namely, $\ve_u,\ \forall u \in \gU$ and $\ve_i,\ \forall i \in \gI_w$. We have $\gC$ to denote the contents of the items, and each item has its respective content features, denoted as $\vc_i$. For the user, we gather the item content list, denoted as $\mC_u$.

\paragraph{\textbf{Restrict Item Cold-start Recommendation.}}
This paper focuses on the most challenging strict cold-start problem, from the view of item cold-start, where the cold items lack any historical behaviors. Under this constraint, the warm items and the cold items lead to two different ways of recommendation. The warm items are recommended with historical user sequences, which are usually encoded into behavior embeddings. Formally, the warm recommendation can be defined as:
\begin{equation}
    \hat{y}^{(w)}_{ui} =
        R(\ve_{u}, \text{Emb}_{cf}(\vs_i), \vc_i), \ i \in \gI_w,
\end{equation}
where $\text{Emb}_{cf}(\cdot)$ denote the collaborative filtering function for behavior embedding.
However, the user sequence set of the cold item is empty, making the cold items recommendation to be organized with the following formula:
\begin{equation}
    \hat{y}^{(c)}_{ui} =
        R(\ve_{u}, \text{Emb}_{cf}(\phi), \vc_i), \ i \in \gI_c.
\end{equation}
Thus the restricted cold-start recommendation problem turns to recommend the above warm items and cold items well.

\section{Methodology}

In this section, we first introduce the general framework of our proposed \model. Next, we assume the existence of a trained LLM simulator and explain how to simulate user sequences using the coupled funnel strategy, which includes coupled filtering simulation and coupled refining simulation. Following this, we elaborate on the training strategy of the LLM and the filter simulator. Lastly, we outline the industrial implementation details and provide a complexity analysis of \model.

\begin{figure*}[t]
\centering
    \includegraphics[width=\linewidth]{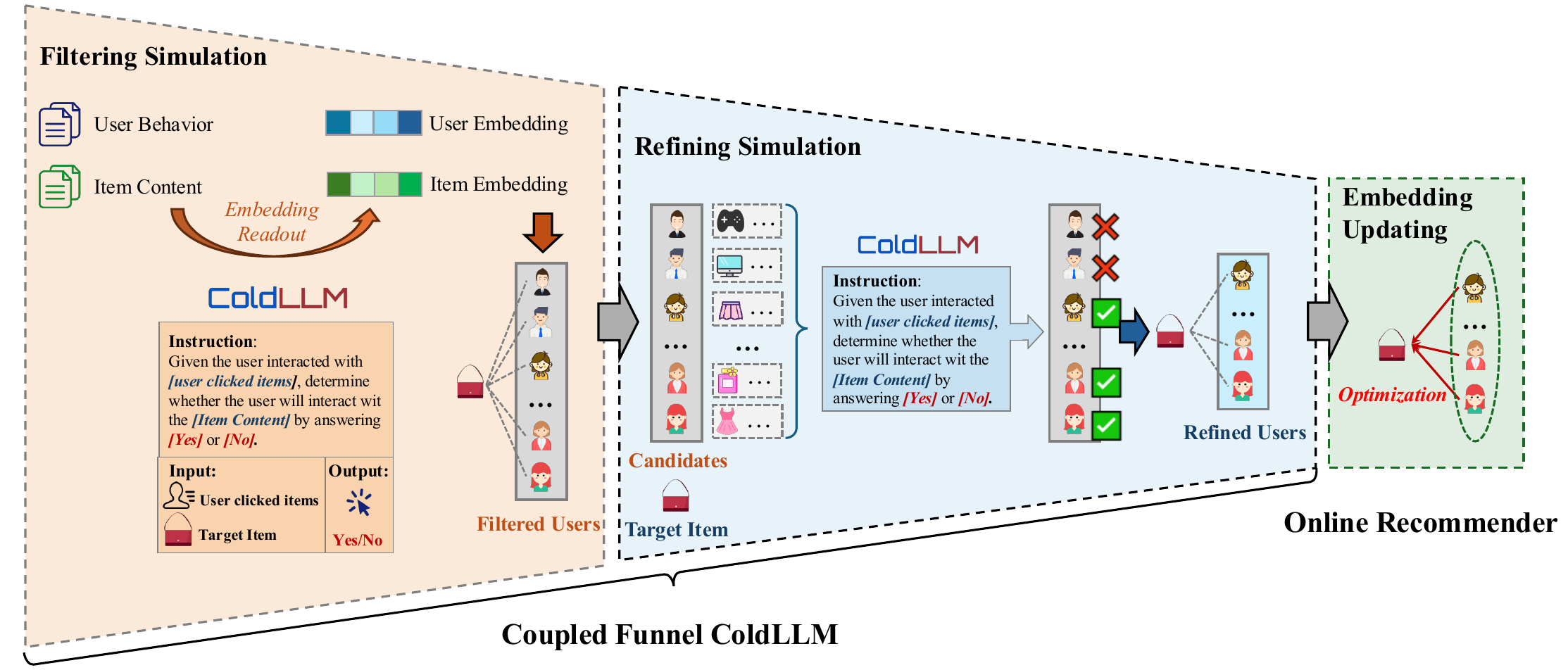}
    \vspace{-1em}
    \caption{The overall model architecture of the proposed \model.}
\label{fig:framework}
\end{figure*}

\subsection{Overall Framework}\label{sec:overall}

The primary distinction between cold and warm items lies in their interaction histories: cold items lack historical behaviors, while warm items have rich interaction data. Traditional models mainly address this cold-start problem through synthetic embedding construction approaches, which can introduce a natural gap between cold and warm items~\cite{chen2022gar,huang2023aldi}. One fundamental solution is to simulate user behaviors for each cold item, and then obtain the cold item embedding from behavior embedding optimization rather than from a mapping function. Building on this concept, we introduce behavior simulation and embedding optimization.
\subsubsection{Behavior Simulation}
The behavior simulation summarizes all the historical behaviors and all the user and item information to simulate possible users that can aid in updating the embeddings for cold items. Considering this, we employ LLM to analyze all the positive historical behaviors to act as a simulator for generating user sequences for cold items. Specifically, the ColdLLM process can be formally defined as follows:
\begin{equation}
    \hat{\vs}_i = \text{ColdLLM}(\vc_i, \gU,\gH,\gC), \forall \ i \in \gI_c.
    \label{eq:ColdLLM}
\end{equation}
In the ideal scenario, the ColdLLM could have access to the information of cold item $\vc_i$, the entire user set $\gU$, the complete historical interactions of all users $\gH$, and the content details of all items $\gC$.

\subsubsection{Embedding Optimization}
By simulating the user sequence for cold items, these items are transformed into warm items. Through simulated behaviors, the recommender system utilizes the existing behavior embedding optimization structure to leverage trained user and warm item embeddings for optimizing the cold item embedding. In offline datasets, such simulation can even enrich the training data to enhance user representation further. For online billion-scale platforms, simulated interactions are used solely to update the cold item embedding. The \textbf{final cold embedding} for downstream tasks can be formally presented as follows:
\begin{equation}
    \ve_i^{(c)} = \text{Emb}_{\text{opt}}(\vc_i, \hat{\vs}_i, \mE),
    \label{eq:emb-opt}
\end{equation}
where $\text{Emb}_{\text{opt}}(\cdot)$ denotes the general behavior embedding optimizer of the recommendation systems, $\ve_i^{(c)}$ represents the embedding of the cold item $i$, and $\vs^{(c)}_i$ is the simulated user sequence for the cold item. $\mE$ denotes all the trained warm embeddings, including the users and the warm items.

\subsection{Coupled Funnel ColdLLM}
Though Eq.~(\ref{eq:ColdLLM}) provides an ideal case of an LLM simulator, compared with traditional embedding-based models, the LLM suffers from heavy computational complexity, requiring more expensive GPUs. This makes the original ColdLLM unsuitable for billion-scale recommendations. In this subsection, we propose the coupled-funnel ColdLLM to incorporate coupled filter models efficiently and effectively simulate cold item behaviors.

\subsubsection{\textbf{Filtering Simulation}}
The aim of the filtering process is to diminish potential users from a dataset of billion-scale to a more manageable range of tens to thousands. Embedding-based filtering models and LLMs encounter distinct challenges in forecasting user behaviors toward cold items. Embedding-based filtering models adaptly embed users and items into vectors but encounter difficulty in capturing users' content-level intent and items' high-level content information. To address this, we initially enhance the filtering models with LLM-processed embeddings, and in the subsequent subsection, we present the coupled training of the filtering model.

Formally, we employ an LLM to extract the content embedding of an item and then apply a matching function to map this embedding for behavior filtering, which can be expressed as follows:
\begin{equation}
\vf_i = \gF_\gI(LLM_{emb}(\vc_i)),
\end{equation}
where $\vf_i$ represents the filtering embedding for item $i$, $\gF_\gI(\cdot)$ is the mapping function, and $LLM_{emb}(\cdot)$ is the LLM embedding readout function.

The embedding readout function is designed to extract the content embedding from the LLM. Specifically, we first obtain the last layer embedding, which represents the processed token information, and then apply mean pooling to derive the content feature embedding for any given cold item:
\begin{equation}
\vf_i = \gF_\gI\left(\frac{1}{|\vc_i|}\sum_{j=1}^{|\vc_i|} \gE^{(L)}(\vc_{i})[j]\right),
\end{equation}
where $\gE^{(L)}(\vc_{i})[j]$ represents the $j$-th embedding of the $L$-th layer of the LLM. Here, $\vc_i$ stands for the content feature of the item, $\vc_{i}[j]$ refers to the $j$-th token in $\vc_{i}$, and $|\vc_i|$ indicates the total number of tokens in $\vc_i$.

To filter users who are likely to interact with the cold item, we consider both content embeddings and behavioral embeddings. We use the dot product of the mapped user embedding and the mapped item embedding to identify the top-$K$ highest score candidates:
\begin{equation}\label{eq:topk}
\vs_i^{(f)} = \text{Top}_K\left(\{\gF_\gU(\ve_u|\mC_u)^\top \cdot \vf_i \mid \forall u \in \gU\}\right),
\end{equation}
where $\gF_\gU$ is the mapping function for the users. Please refer to the following subsection for design and training details. 

This top-$K$ computation can be accelerated using efficient similarity search platforms, such as FAISS~\cite{johnson2019faiss}, resulting in $\gO(1)$ computational complexity.

\subsubsection{Refining Simulation}
After filtering, the candidate user pool will decrease from billions to tens or hundreds. Next, we employ LLMs for examination and enhancement. For each iteration, we feed the user's context and the item's content into the LLM, which outputs a prediction of whether the user will interact with the item, displaying 'yes' or 'no'.
The refining module considers three technique details aspects:  1. Context construction, 2. Prompt Design, and 3. Refining Process.
\paragraph{\textbf{Context Construction}}
LLMs rely on users's context to judge whether the user will interact with the recommendation. However, the users may have too many historical behaviors, or not all the historical behaviors are correlated to the query item. To this end, we utilize the item embedding from the filtering process to filter the related items,
\begin{equation}
    \mC_u^{(f)} = \text{Top}_L\left(\{\vf_i^\top \cdot \vf_j \mid \forall j \in \mC_u\}\right),
\end{equation}
where $\mC_u$ denotes the user's historically interacted items.

\paragraph{\textbf{Prompt Design}}
The LLM prompt contains three parts: 1. fixed prompts, 2. user context, and 3. item content. The fixed prompts are used to set up the goal of the LLM Simulator. As shown in Figure~\ref{fig:framework}, the fixed prompt is 
\begin{quote}

``Given the user interacted with {\textcolor{blue}{[\text{Text}$(\mC_u^{(f)})$]}}, determine whether the user will interacted the {\textcolor{blue}{[\text{Text}$(\vc_i)$]}} by answering {\color{red}Yes} or {\color{red}No}.''
\end{quote}

Then given an LLM, the pair-wise LLM simulation can be present as follows:
\begin{equation}
    \hat{Z}_{u,i} = \left\{
        \renewcommand{\arraystretch}{1.5}{
                \begin{array}{lcl}
                    1, & \text{LLM}(\mC_u^{(f)}, \vc_i) = "Yes" \\
                    0, & \text{LLM}(\mC_u^{(f)}, \vc_i) = "No"
                \end{array}
                }
            \right. ,
\end{equation}
where $\hat{Z}_{u,i}$ represents the predicted value of the LLM. Subsequent research could consider employing a more intricate framework to obtain a continuous value.

\paragraph{\textbf{Refine Process}} During the refining process, we traverse the filtered user set and only maintain the users that are predicted as ``yes'' by the LLM simulator. Thus finally, the simulated users can be obtained as follows,
\begin{equation}
\vs_i^{(r)} = \{u|\hat{Z}_{u,i}=1 \mid \forall u \in \vs_i^{(f)}\}.
\end{equation}

After obtaining the refined user simulation results, the sequence can be fed into the behavior embedding optimization framework~(Eq.~(\ref{eq:emb-opt})) to enhance the cold item embedding.

\subsection{Simulator Training}\label{sec:training}
In this subsection, we introduce the simulator training, including the training of the based LLM model and the filtering model.
\subsubsection{LLM Training}
To better fit the recommendation data of each different recommendation scenario, we utilize a Low-Rank fine-tuning strategy to ensure the LLM can capture the data distribution of the recommendation scenario, and capture the trends by performing online model updating.

\paragraph{\textbf{Data Preparation}}
We use user item behaviors to train our LLM. In the online recommendation, we encounter three types of behaviors: positive behaviors (users click on an item), negative behaviors (users ignore the item), and unobserved behavior (the recommender system does not recommend the item to the user, and we cannot observe the user's intention).

As negative behaviors make up approximately 70\%-90\% of online behaviors, we perform a 1:1 sampling of positive and negative behaviors to address the imbalanced distribution. Additionally, we conduct an extra 1:1 sampling of positive and unobserved behavior to ensure that the LLM can capture a broader range of intentions, especially for cold items.
For offline datasets, we only use a 1:1 sampling of positive and unobserved data to train the LLM.

\paragraph{\textbf{Fine-tuning Structure}}
Specifically, we add an additional low-rank modification matrix on all the parameters of the transformer structure, in terms of Q, K, and V in the self-attention mechanism, as well as the feed-forward layers. This Low-Rank Adaptation (LoRA)~\cite{hu2021lora} approach allows us to efficiently fine-tune the large language model for specific recommendation scenarios while maintaining most of the pre-trained weights.
Specifically, we add trainable rank decomposition matrices to each weight matrix of the original model. For a given weight matrix $W \in \mathbb{R}^{d \times g}$, we add a low-rank update:
\begin{equation}
W' = W + \Delta W = W + BA,
\end{equation}
where $B \in \mathbb{R}^{d \times r}$ and $A \in \mathbb{R}^{r \times g}$ are the low-rank decomposition matrices, with $r$ representing the rank of the decomposition, which is generally much smaller than the dimensions $d$ and $g$.

The update on the transformer networks can be written as follows:
\begin{align}
Q &= W_Q x + \Delta W_Q x = W_Q x + B_Q A_Q x, \\
K &= W_K x + \Delta W_K x = W_K x + B_K A_K x, \\
V &= W_V x + \Delta W_V x = W_V x + B_V A_V x,
\end{align}

where $W_Q$, $W_K$, and $W_V$ are the original weight matrices for Query, Key, and Value projections,
$B_Q$, $B_K$, $B_V$ and $A_Q$, $A_K$, $A_V$ are the low-rank decomposition matrices for each projection.

Similarly, for the feed-forward layers, we have:
\begin{equation}
\text{FFN}_{\text{LoRA}}(x) = \text{ReLU}(x(W_1 + A_1B_1) + b_1)(W_2 + A_2B_2) + b_2.
\end{equation}
where $B_1$, $A_1$, $B_2$, and $A_2$ are the low-rank decomposition matrices for the FFN.

\subsubsection{Coupled Filter Model Training}
The coupled filter model has two design proposals: 1. reflect user item behaviors; 2. coupled with the LLM. Specifically, we utilize the combination of two pairs of embedding to accomplish this purpose.

\paragraph{\textbf{Training of the Behavior Filtering}}
For every given user-item pair $(u,i)$, a negative pair $(u,j)$ is randomly selected. These pairs can be collectively represented as a triple $(u,i,j)$. The output of behavior filtering can be expressed as $\hat{Y}_{ui}^{(B)} = \gF_\gU^{(B)}(\ve_u)^\top \cdot \gF_\gI^{(B)}(LLM_{emb}(\vc_i))$.
We consider BPR loss~\cite{rendle2009bprmf} to optimize the recommendation performance of the behavior filtering model 
\begin{equation}
\mathcal{L}_{BPR} = -\sum_{(u,i,j)} \ln \sigma(\hat{Y}_{ui}^{(B)} - \hat{Y}_{uj}^{(B)}),
\end{equation}
where $\sigma(\cdot)$ is the sigmoid function. This loss encourages the filter model to rank the positive item higher than the negative item. Besides, we also utilize the aligning loss in ALDI~\cite{huang2023aldi} to help the training of behavior filtering.

\paragraph{\textbf{Training of the Coupled ColdLLM Filtering}}
For the coupled LLM filtering, we apply the formula below to filter the users: $\hat{Y}_{ui}^{(L)} = \mF_{\mU}^{(L)}(\ve_u)^\top \cdot \mF_{\mI}^{(L)}(LLM_{emb}(\vc_i))$ In addition to the BPR loss, we introduce the coupled ColdLLM loss to maintain similarity with the ColdLLM in the coupled filter model:
\begin{equation}
    \label{eq:ranking_gap}
    \gL_{coupled} = -\sum_{(u,i)} \left(\hat{Z}_{ui} \ln \hat{Y}_{ui}^{(L)} + (1 - \hat{Z}_{ui}) \ln(1 - \hat{Y}_{ui}^{(L)}) \right).
\end{equation}

\subsection{Implementation Strategy}\label{sec:online}

ColdLLM has demonstrated its effectiveness by providing helpful cold-start recommendations on the e-commerce platform.
In this subsection, we detail the online deployment of ColdLLM and analyze its computational complexity.

\begin{figure}[tb]
    \centering
    \includegraphics[width=\linewidth, trim=0cm 0cm 0cm 0cm,clip]{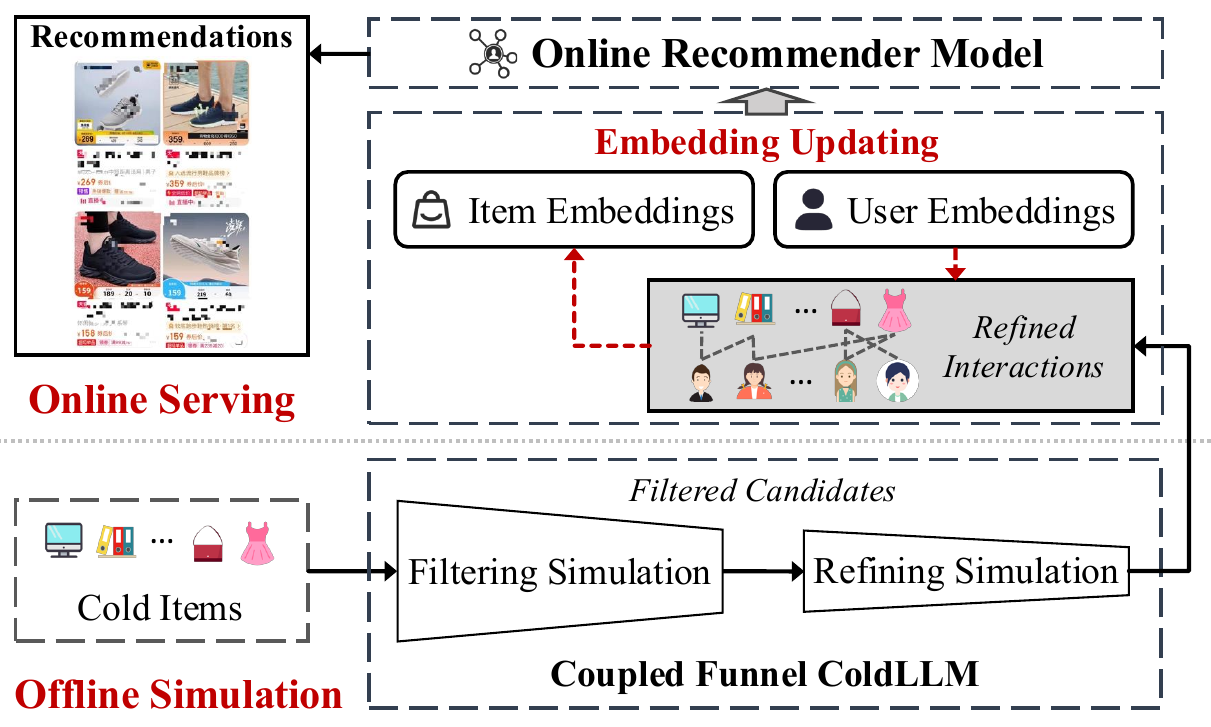}
    \vspace{-1em}
    \caption{The system architecture for ColdLLM deployment.}
    \vspace{-1em}
    \label{fig:onlineArch}
\end{figure}

\paragraph{\textbf{Real-World Deployment}}
As shown in Fig.~\ref{fig:onlineArch}, our overall framework consists of three primary components: (i) online serving; (ii) online training (embedding updating); and (iii) offline simulation.

When new items are uploaded to the platform, we first employ our model to simulate user interactions for embedding updates. These simulated user-item pairs are then fed into the online embedding updating structure. Since these interactions are simulated rather than actual user behaviors, we update only the embeddings of the cold items. Finally, we transmit the updated cold item embeddings to the online recommendation service.

\paragraph{\textbf{Complexity Analysis}}
The computational complexity of ColdLLM comprises three main components: coupled filtering complexity, coupled refining complexity, and embedding update complexity.

\begin{enumerate}[leftmargin=*]
    \item \textbf{Coupled Filtering:} Leveraging similarity indexing frameworks like FAISS, we can efficiently downscale the candidate users from billions to hundreds with a complexity of $\mathcal{O}(1)$ in approximately 60 ms.

    \item \textbf{Coupled Refining:} We refine the filtered candidates using fine-tuned LLaMA-7B models to identify 20 qualified users. This process takes about 200-400 ms for each user-item pair. In total, the LLM-refining stage requires less than 8 seconds.

    \item \textbf{Embedding Update:} The online embedding process utilizes the simulated interactions to optimize the cold item embeddings within 120 ms.
\end{enumerate}
To enhance efficiency, we employ parallel processing techniques. When equipped with three 8$\times$A100 GPU machines, our system can handle a load of 8,640 cold items per hour. This capacity can be easily scaled by adding more GPUs to the infrastructure.

\section{Experiments}
In this section, we conduct comprehensive experiments on benchmark cold-start recommendation datasets, aiming to answer the following research questions.
\textbf{RQ1:} Does \model outperform contemporary state-of-the-art cold-start recommendation models in overall, warm, and cold recommendations?
\textbf{RQ2:} What is the effect of different components in \model?
\textbf{RQ3:} How do key hyper-parameters impact the performance of \model?
\textbf{RQ4:} How does \model perform in real-world industrial recommendations?

\begin{table*}[h]
  \centering
  \caption{Comparison results on overall, cold, and warm item recommendations over three backbone models (MF, NGCF, LightGCN). The best and second-best results in each column are highlighted in \textbf{bold} font and \underline{underlined}.}
  \resizebox{0.945\linewidth}{!}{
    \begin{tabular}{cc|cc|cc|cc|cc|cc|cc}
    \toprule
    \multicolumn{2}{c|}{\multirow{3}{*}{Method}} & \multicolumn{4}{c|}{Overall Recommendation} & \multicolumn{4}{c|}{Cold Recommendation} & \multicolumn{4}{c}{Warm Recommendation} \\
    \multicolumn{2}{c|}{} & \multicolumn{2}{c|}{CiteULike} & \multicolumn{2}{c|}{MovieLens} & \multicolumn{2}{c|}{CiteULike} & \multicolumn{2}{c|}{MovieLens} & \multicolumn{2}{c|}{CiteULike} & \multicolumn{2}{c}{MovieLens} \\
    \multicolumn{2}{c|}{} & Recall & NDCG  & Recall & NDCG  & Recall & NDCG  & Recall & NDCG  & Recall & NDCG  & Recall & NDCG \\
    \midrule
    \multirow{10}{*}{\rotatebox{90}{MF}} 
    & DropoutNet & 0.0794& 0.0670& 0.0646& 0.1127& 0.2268& 0.1356& 0.0671& 0.0808& 0.1343& 0.0792& 0.1391& 0.1386\\
    & MTPR  & 0.1060& 0.0810& 0.0739& 0.1055& 0.2496& 0.1476& \underline{0.0745}& \underline{0.0811}& 0.1728& 0.0998& 0.1628& 0.1388\\
    & CLCRec & 0.1269& 0.0992& 0.0784& 0.1358& 0.2295& 0.1347& 0.0744& 0.0726& 0.1898& 0.1167& 0.1699& 0.1696\\
    \cmidrule{2-14} 
	& DeepMusic & 0.0956& 0.0789& \underline{0.0933}& \underline{0.1377} & 0.2141& 0.1262& 0.0327& 0.0391& \underline{0.2838} & \underline{0.1933} & \underline{0.2076} & \underline{0.1819} \\
    & MetaEmb & 0.0972& 0.0804& \underline{0.0933}& \underline{0.1377} & 0.2232& 0.1306& 0.0432& 0.0468& \underline{0.2838} & \underline{0.1933} & \underline{0.2076} & \underline{0.1819}\\
 & GNP & 0.1568& \underline{0.1305}& 0.0813& 0.1253& 0.2107& 0.1193& 0.0683& 0.0704& \underline{0.2838} & \underline{0.1993} & \underline{0.2076} & \underline{0.1819}\\
    & GAR   & 0.1440& 0.1132& 0.0462& 0.0812& 0.2453& 0.1479& 0.0348& 0.0510& 0.2272& 0.1438& 0.1003& 0.1003\\
    & ALDI & \underline{0.1618}& 0.1204& 0.0914& 0.1355& \underline{0.2684}& \underline{0.1550}& 0.0431& 0.0464& \underline{0.2838}& \underline{0.1993}& \underline{0.2076}& \underline{0.1819}\\
	\cmidrule{2-14} 
	& \textbf{\model}  & \textbf{0.2213}& \textbf{0.1599}& \textbf{0.0976}& \textbf{0.1473}& \textbf{0.3218}& \textbf{0.1875}& \textbf{0.1225}& \textbf{0.1182}& \textbf{0.3058}& \textbf{0.2061}& \textbf{0.2129}& \textbf{0.1924}\\
    & \textit{Improvement} & 36.77\%& 22.52\%& 4.61\%& 6.98\%& 19.90\%& 20.96\%& 64.43\%& 45.75\%& 7.75\%& 3.41\%& 2.56\%& 5.78\%\\
    \midrule
    \midrule
    \multirow{12}{*}{\rotatebox{90}{NGCF}} 
    & DropoutNet & 0.0813& 0.0656& 0.1144& 0.1935& 0.2211& 0.1278  & 0.0214  & 0.0223  & 0.1416  & 0.0842& 0.2517  &0.2457 \\
    & MTPR  & 0.1006  & 0.0769  & 0.1132& 0.1818& 0.2479& 0.1391  & 0.0749  & 0.0894  & 0.1753  & 0.0977  & 0.2509  & 0.2329  \\
    & CLCRec & 0.1201  & 0.0920  & 0.1270  & 0.2042  & 0.2093  & 0.1188  & 0.0694  & 0.0777  & 0.1886  & 0.1136  & 0.2807  & 0.2627  \\
	\cmidrule{2-14} 
	& DeepMusic & 0.1269  & 0.1043  & 0.1393  & 0.2265  & 0.1980  & 0.1152  & 0.0409  & 0.0493  & 0.2347& 0.1485& 0.3076 & 0.2922 \\
    & MetaEmb & 0.1119  & 0.0957  & 0.1393  & 0.2265 & \underline{0.2830}  & \underline{0.1664}  & 0.0247  & 0.0245  & 0.2347&0.1485& 0.3076 & 0.2922 \\
    & GNP & 0.1449  & 0.1099  & 0.1381  & 0.2234  & 0.2318  & 0.1371  & 0.0589  & 0.0646  & 0.2347& 0.1485& 0.3076  & 0.2922  \\
    & GAR   & 0.1144  & 0.0909  &0.0098  & 0.0174  & 0.2099  & 0.1251  & 0.0166  & 0.0178  & 0.1850  & 0.1143  & 0.2788  & 0.2721  \\
        & ALDI & \underline{0.1541}  & 0.1141  & 0.1393  & 0.2206  & 0.2466  & 0.1399  & \underline{0.1022}  & \underline{0.1113}  & 0.2347& 0.1485& 0.3076  & 0.2922  \\
        \cmidrule{2-14} 
        & UCC & 0.1094  & 0.0990  & 0.0981  & 0.1523  & 0.0019  & 0.0008  & 0.0063  & 0.0066  & 0.2402 & 0.1582 & 0.2191  & 0.1999  \\
        & MI-GCN &  0.1324 & \underline{0.1158} & \underline{0.1401}  & \underline{0.2299} & 0.0641 & 0.0335 & 0.0162 & 0.0187 & \underline{0.2827} & \underline{0.1815} & \underline{0.3104} & \underline{0.2973} \\
	\cmidrule{2-14} 
	& \textbf{\model}  & \textbf{0.2211} & \textbf{0.1653} & \textbf{0.1447}& \textbf{0.2372}& \textbf{0.3404} & \textbf{0.1986} & \textbf{0.1670}& \textbf{0.1662}& \textbf{0.2992} & \textbf{0.1958} & \textbf{0.3186}& \textbf{0.3052}\\
    & \textit{Improvement} & 43.48\% & 42.75\% & 3.28\%& 3.18\%& 20.28\% & 19.35\% & 63.41\%& 49.33\%& 5.84\%& 7.88\%& 2.64\%& 2.66\%\\
    \midrule
    \midrule
    \multirow{12}{*}{\rotatebox{90}{LightGCN}} & 
    DropoutNet & 0.0883  & 0.0639  & 0.1165  & 0.1978  & 0.2309  & 0.1312  & 0.0340  & 0.0373  & 0.1175  & 0.0692  & 0.2560  & 0.2518  \\
    & MTPR  & 0.1001  & 0.0753  & 0.1011  & .0.1551  & 0.2585& 0.1454  & 0.0779  & 0.0802  & 0.1753  & 0.0697  & 0.2247  & 0.2009  \\
    & CLCRec & 0.1293  & 0.0965  & 0.1253  & 0.2037  & 0.2435  & 0.1425  & 0.0677  & 0.0816  & 0.2149  & 0.1302  & 0.2764  & 0.2612  \\
	\cmidrule{2-14} 
	& DeepMusic & 0.0985  & 0.0745  & 0.1418  & 0.2330 & 0.2239  & 0.1259  & 0.0635  & 0.0719  & 0.2528& 0.1541& 0.3130& 0.3008 \\
    & MetaEmb & 0.0924  & 0.0714  & 0.1418  & 0.2330& 0.2252  & 0.1295  & 0.0248  & 0.0244  & 0.2528& 0.1541& 0.3130 & 0.3008 \\
    & GNP & 0.1609  & 0.1197  & 0.1293  & 0.2106  & 0.2606  & 0.1532  & 0.0771& 0.0760  & 0.2528  & 0.1541  & 0.3130 & 0.3008 \\
    & GAR   & 0.1357& 0.1062& 0.0106& 0.0195& 0.2539  & 0.1489& 0.0110  & 0.0130  & 0.2339  & 0.1455  & 0.2873  & 0.2794  \\
     & ALDI   & \underline{0.1626}& 0.1201     & 0.1428 & 0.2316  & \underline{0.2692}& \underline{0.1539}& \underline{0.1229}& \underline{0.1295}  & 0.2528     & 0.1541     & 0.3130 & 0.3008 \\
	\cmidrule{2-14} 
 & UCC & 0.1374& 0.1260& 0.1277  & 0.2020  & 0.0020& 0.0011& 0.0063  & 0.0073  & 0.3002& 0.2010 & 0.2830  & 0.2641  \\
 & MI-GCN & 0.1557 & \underline{0.1483} &  \underline{0.1454} & \underline{0.2378}  & 0.0507 & 0.0262 & 0.0307 & 0.0312  & \textbf{0.3372} & \textbf{0.2362}  &  \underline{0.3227} & \underline{0.3068} \\
 \cmidrule{2-14} 
	& \textbf{\model} & \textbf{0.2285} & \textbf{0.1747} & \textbf{0.1506}& \textbf{0.2468}& \textbf{0.3601} & \textbf{0.2126} & \textbf{0.1759}& \textbf{0.1762}& \underline{0.3252} & \underline{0.2156} & \textbf{0.3314}& \textbf{0.3186}\\
    & \textit{Improvement} & 40.52\%& 17.80\%& 3.58\%& 3.78\%& 33.76\%& 38.14\%& 43.12\%& 36.06\%& -- & -- & 2.70\%& 3.85\%\\
    \bottomrule
    \end{tabular}%
    }
  \label{tab:mainresult}%
\end{table*}%

\subsection{Experimental Setup}
\subsubsection{Datasets}
We conduct experiments on two widely used datasets: \textbf{CiteULike}\footnote{\url{https://github.com/js05212/citeulike-a}}~\cite{wang2013citeulike}, containing 5,551 users, 16,980 articles, and 204,986 interactions, and \textbf{MovieLens}\footnote{\url{https://grouplens.org/datasets/movielens/10m}}~\cite{harper2015movielens}, comprising 6,040 users, 3,883 items, and 1,000,210 interactions. 
For each dataset, following previous works~\cite{huang2023aldi}, 20\% items are designated as cold-start items, with interactions split into a cold validation set and testing set (1:1 ratio). Records of the remaining 80\% of items are divided into training, validation, and testing sets, using an 8:1:1 ratio.

\subsubsection{Compared Baselines}
To assess the effectiveness of our proposed \model, we conducted a comprehensive analysis with ten leading-edge models in the domain of cold-start item recommendations, which can be categorized into three main groups.
(i) Dropout-based embedding simulation models: \textbf{DropoutNet}~\cite{volkovs2017dropoutnet}, \textbf{MTPR}~\cite{du2020mtpr}, and \textbf{CLCRec}~\cite{wei2021clcrec}. (ii) Generative-based embedding simulation models: \textbf{DeepMusic}~\cite{van2013deepmusic}, \textbf{MetaEmb}~\cite{pan2019metaemb}, \textbf{GNP}~\cite{chen2025graph}, \textbf{GAR}~\cite{chen2022gar}, and \textbf{ALDI}~\cite{huang2023aldi}. (iii) User behavior simulation models: \textbf{UCC}~\cite{liu2023ucc} and \textbf{MI-GCN}~\cite{wang2024mutual}. To further verify the universality of \model, we verify these models on three widely used recommendation backbones: \textbf{MF}~\cite{rendle2009bprmf}, \textbf{NGCF}~\cite{wang2019ngcf}, and \textbf{LightGCN}~\cite{he2020lightgcn}.

\subsubsection{Hyperparameter Setting}
In the filtering phase, we utilized AdamW as the optimizer with a chosen learning rate of $1\times10^{-5}$, and set the batch size for each training batch to 128. We opted for a top-k value of 20. In the refining phase, the learning rate was adjusted to $5\times10^{-5}$. 
The dimension of the embeddings was standardized to 200 for all models. We employed the Adam optimizer with a learning rate of $1\times10^{-3}$ and applied early stopping by monitoring NDCG on the validation set.

\subsubsection{Evaluation Metrics}
Our evaluation encompasses the overall, warm, and cold recommendation performance, adopting a widely adopted full-ranking evaluation approach~\cite{he2020lightgcn,huang2023aldi}. 
Specifically, we employ Recall@$K$ and NDCG@$K$ as our primary metrics, where $k=20$. Following previous works~\cite{chen2022gar,huang2023aldi}, during the testing, we randomly select 2,000 users for evaluation.

\subsection{Main Results (RQ1)}
The performance comparison of overall, warm, and cold recommendations between \model and other baselines on benchmark datasets is presented in \autoref{tab:mainresult}. 
From the results, we can have the following observations:

\textbf{\model can achieve significant improvements over current methods.} From the table, we can find that \model can consistently demonstrate superiority across different datasets and backbones. Specifically, \model brings an average NDCG improvement of 10.79\% and 37.10\% on overall and cold item recommendations over LightGCN. This enhancement illustrates the effectiveness of \model with the coupled-funnel behavior simulation based on the LLM's world knowledge.

\textbf{The generative-based embedding simulation models generally perform better in warm and overall recommendations than dropout-based embedding simulation models.} This indicates that forcing the warm behavior embeddings and cold content embeddings to align with each other through the same embedding layer may lead to a performance drop in the warm item recommendation. The interaction simulation with \model addresses this by allowing cold and warm items to be adequately trained within a unified recommender.

\textbf{Existing behavior simulation models retain relatively good performance in overall and warm recommendations, but fall short in cold recommendations.} A possible reason is that the behavior generation only based on the content information with DNNs is insufficient for accurate behavior simulation for cold items. \model leverages the world knowledge of LLMs to utilize more information for the behavior simulation of cold items.

\vspace{-1em}
\subsection{Ablation Study (RQ2)}
\begin{figure}[t]
\centering
    \includegraphics[width=0.95\linewidth]{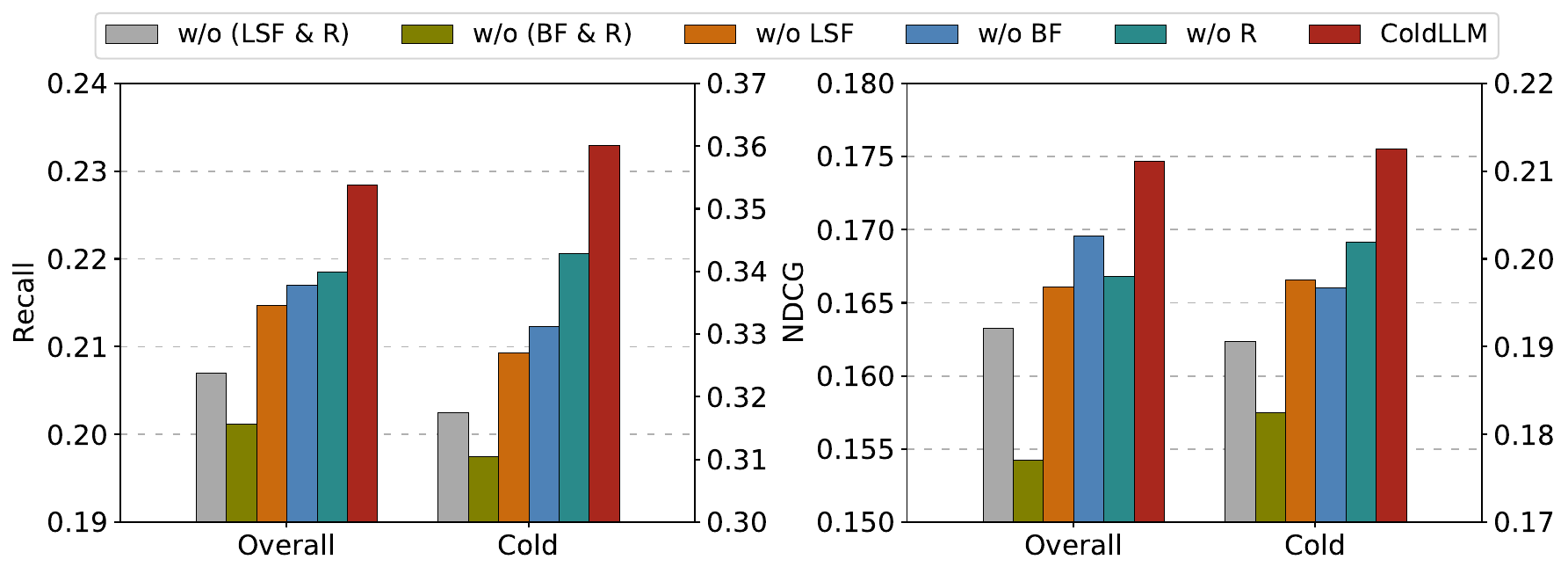}
    \vspace{-1em}
    \caption{Ablation study results on CiteULike.}
    \vspace{-0.8em}
\label{fig:ablation}
\end{figure}

We conduct an ablation study of our proposed \model approach to validate its key components, of which the results are illustrated in Figure~\ref{fig:ablation}. Specifically, we compare \model with its five variants: 
(i) \textbf{\textit{w/o LSF \& R}} removes the coupled ColdLLM filtering and the refining simulation.
(ii) \textbf{\textit{w/o BF \& R}} removes the behavior filtering and the refining simulation.
(iii) \textbf{\textit{w/o LSF}} skips the coupled ColdLLM filtering module.
(iv) \textbf{\textit{w/o BF}} skips the behavior filtering module.
(v) \textbf{\textit{w/o R}} skips the refining simulation. 
Further, we compare the adoption rate of the filtered users by the filtering simulation of \model with three different strategies, as shown in Table~\ref{tab:ar}.
From the results, we can have the following observations:

\textbf{Effectiveness of the filtering simulation}. The decline in performance for \textit{w/o LSF} and \textit{w/o BF} demonstrates the effectiveness of the filtering simulation. Further, the more pronounced drop in performance for \textit{w/o LSF} highlights the importance of world knowledge in LLMs for generating behavior patterns. Additionally, the adoption rates in Table~\ref{tab:ar} indicate that the filtering simulation yields behaviors of superior quality compared to alternative strategies.

\textbf{Effectiveness of refining stage}. 
The necessity of the refining stage is evident when comparing the performance with \textit{w/o R}. Furthermore, the model \textit{w/o LSF \& R} and \textit{w/o BF \& R}, exhibit a more significant performance decline than \textit{w/o LSF} and \textit{w/o BF}. This contrast also underscores the meaningful contribution of the refining simulation to the overall effectiveness of \model.

\begin{table}[tbp]
  \centering
  \caption{Adoption rate of different filtering strategies.}
  \vspace{-1em}
  \resizebox{0.9\linewidth}{!}{
    \begin{tabular}{c|ccc|c}
    \toprule
    Strategy & Random	 & w/o LSF & w/o BF & \textbf{\model} \\
    \midrule
    Adoption rate (\%) & 8.18\%	& 80.45\% & 81.33\%  & \textbf{85.96\%} \\
    \bottomrule
    \end{tabular}%
    \vspace{-1.2em}
   }
  \label{tab:ar}%
\end{table}%

\subsection{Parameter Study (RQ3)}
\begin{figure}[t]
\centering
    \includegraphics[width=0.975\linewidth]{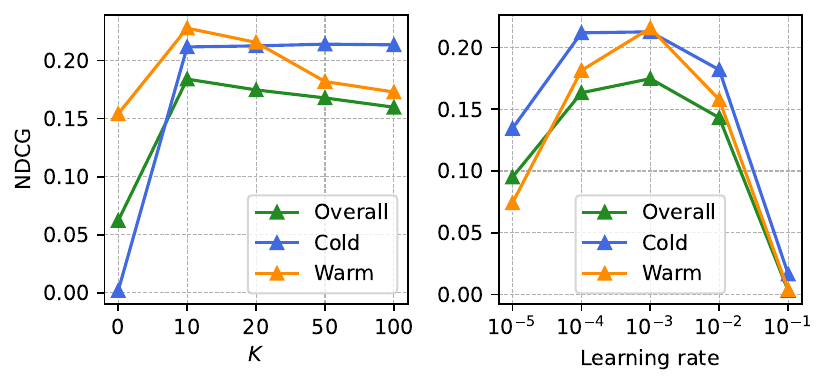}
    \vspace{-1em}
    \caption{Parameter study results on CiteULike.}
    \vspace{-1em}
\label{fig:param}
\end{figure}

In this subsection, we study the impact of key
hyper-parameters on \model with CiteULike, including the filtering candidate number $K$ in Eq.(\ref{eq:topk}) and the learning rate of embedding updating. The results are shown in Figure~\ref{fig:param}.

\textbf{Effect of filtering candidate number $K$}. From the results, we can find that optimal results for overall and warm recommendations are achieved with a modest value of $K$, such as $K$=10 for CiteULike. Conversely, a larger $K$ is beneficial for cold recommendations, with $K$=50 for CiteULike yielding the best outcomes. However, an excessively large $K$ can degrade performance by introducing noise from irrelevant interactions.

\textbf{Effect of the updating learning rate}. From the figure, we can observe that the three types of recommendation tasks achieve the best results with a similar optimal learning rate, which indicates that the tuning of the learning rate would simultaneously be suitable for all three tasks.

\subsection{Online Evaluation (RQ4)}

To validate the effectiveness of \model in an industrial setting, we conducted an online A/B test on one of the largest e-commerce platforms. The experiment ran for two consecutive weeks, involving 5\% of users for each group. We compared \model against three baselines: \textbf{Random}, \textbf{MetaEmb}~\cite{pan2019metaemb}, and \textbf{ALDI}~\cite{huang2023aldi}. Table~\ref{tab:ab_test} presents the results of these online A/B tests.

\paragraph{\textbf{Evaluation Metrics}}
We employed three tailored metrics to assess the performance of ColdLLM against existing baselines:

\begin{itemize}[leftmargin=*]
    \item Page Views (Cold-PV): The number of user clicks during the cold-start period.
    \item Page Click-Through Rate (Cold-PCTR): The ratio of clicks to impressions during the cold-start period.
    \item Gross Merchandise Value (Cold-GMV): The total value of user purchases during the cold-start period.
\end{itemize}

We define the cold-start period as the interval from the item's publication to two hours after its release.

\paragraph{\textbf{Results and Analysis}}
The results in Table~\ref{tab:ab_test} demonstrate that \model consistently outperforms both baselines across all metrics. Specifically, compared to the random baseline, \model achieves substantial improvements of 11.45\% in Cold-PV, 5.60\% in Cold-PCTR, and 23.80\% in Cold-GMV for cold items. When compared to MetaEmb, \model shows significant gains of 9.20\% in Cold-PV, 4.35\% in Cold-PCTR, and 18.25\% in Cold-GMV. Even against the strong ALDI baseline, \model maintains superior performance with impressive improvements of 7.25\% in Cold-PV, 3.70\% in Cold-PCTR, and 16.90\% in Cold-GMV.

These remarkable improvements across all metrics underscore the effectiveness of \model in addressing the item cold-start problem in real-world recommender systems. The consistent and substantial performance gains, particularly in Cold-GMV, highlight the practical impact of our approach on business outcomes in e-commerce settings.

\begin{table}[h]
\centering
\caption{Results of online A/B tests.}
\vspace{-1em}
\resizebox{0.85\linewidth}{!}{
\begin{tabular}{lccc}
\hline
\textbf{A/B Test} & \textbf{Cold-PV} & \textbf{Cold-PCTR} & \textbf{Cold-GMV} \\
\hline
 vs. Random       & +11.45\%             & +5.60\%                & +23.80\%              \\
vs. MetaEmb      & +9.20\%              & +4.35\%                & +18.25\%              \\
vs. ALDI         & +7.25\%              & +3.70\%                & +16.90\%              \\
\hline
\end{tabular}}
\vspace{-1em}
 \label{tab:ab_test}%
\end{table}

\section{Conclusion}
In this paper, aiming to address current limitations, we propose ColdLLM, which fundamentally solves the cold-start problem in large-scale recommendation systems, significantly improving performance and economic impact. Both online and offline experiments verified the strength of our proposed ColdLLM. Based on the observation, ColdLLM opens new possibilities for leveraging large language models in large-scale online recommendations.

%%
%% The acknowledgments section is defined using the "acks" environment
%% (and NOT an unnumbered section). This ensures the proper
%% identification of the section in the article metadata, and the
%% consistent spelling of the heading.
\begin{acks}
This work was supported in part by the National Natural Science Foundation of China (No. 62272200, No. 62172443) and Hunan Provincial Natural Science Foundation of China (No. 2022JJ30053).
\end{acks}

\section*{Ethical Considerations}
Our paper focuses on leveraging LLMs to simulate the behaviors of cold items. We believe that pursuing this direction is essential, as it not only reveals the untapped potential of LLMs for a large amount of newly occurred items in recommender systems, while concurrently exhibiting the potential ethical considerations. Here we list two main considerations as follows:

(1) \textbf{Industrial Scenario}: We discuss the application in an industrial scenario, which will benefit industrial users and inspire communication between industrial researchers and academic researchers. Nevertheless, it may also pose computational and storage challenges for academic researchers.

(2) \textbf{LLM Resource Cost}: LLMs' training and deployment require significant computational power, potentially leading to higher energy consumption and environmental impact. Additionally, a minimum of one GTX 3090 GPU may be required for the experiments.

%%
%% The next two lines define the bibliography style to be used, and
%% the bibliography file.
\bibliographystyle{ACM-Reference-Format}
\bibliography{reference}

%%
%% If your work has an appendix, this is the place to put it.
%\appendix

\end{document}